\documentclass[11pt,twocolumn]{article}
\usepackage{amsmath}
\usepackage{amsfonts}
\usepackage{graphicx}
\usepackage[super,articletitle=true]{achemso}
\usepackage{hyperref}
\usepackage{geometry}
\usepackage{titlesec}
\usepackage{fancyhdr}
\usepackage{xcolor}
\usepackage{subcaption}
\usepackage{hyperref}
\usepackage{xr-hyper}
\usepackage{soul}
\usepackage[font=small]{caption}
\makeatletter
\renewcommand{\thefootnote}
\makeatother
\titleformat{\section}{\large\bfseries}{\thesection.}{1em}{}
\titleformat{\subsection}{\normalsize\bfseries}{\thesubsection.}{1em}{}
\title{\Large \textbf{Reticulating Local Dynamics in Three-Dimensional Covalent Organic Frameworks}}
\author{
    \begin{minipage}[t]{\textwidth}
    \centering
{Francesco Tavani}\textsuperscript{1,2,3}, Saber Mirzaei\textsuperscript{1}, M. Saeed Mirzaei\textsuperscript{1,4}, Jian Yin\textsuperscript{1}, Yen-hsu Lin\textsuperscript{1}, Caden Myers\textsuperscript{5}, Cheng-Hung Lin\textsuperscript{6}, AM Milinda Abeykoon\textsuperscript{6},
Simon J. L. Billinge\textsuperscript{7}, Omar M. Yaghi\textsuperscript{1,8,\color{blue}†}\\
    \vspace{0.5cm}
    \textsuperscript{1}Department of Chemistry, University of California, Berkeley, CA, United States \\
    \textsuperscript{2}Bakar Institute of Digital Materials for the Planet, Division of Computing, Data Science, and Society, University of California, Berkeley, CA, USA \\
    \textsuperscript{3}Dipartimento di Chimica, Università degli Studi di Roma La Sapienza,\\ P.le A. Moro 5, I-00185, Rome, Italy \\
    \textsuperscript{4}Department of Food Science, College of Agriculture and Life Sciences, Cornell University, Ithaca, New York 14853, United States \\
    \textsuperscript{5}Department of Applied Physics and Applied Mathematics, Columbia University, New York 10027, USA\\
    \textsuperscript{6}National Synchrotron Light Source II, Brookhaven National Laboratory, Upton, \\New York 11973, USA\\
    \textsuperscript{7}Department of Materials, University of California Santa Barbara, Santa Barbara, California 93106, United States\\ 
    \textsuperscript{8}Current Address: Departments of Chemistry and Chemical Engineering, and AIMATRY, Tsinghua University, Beijing, China\\  
    \vspace{0.5cm}
    Corresponding author: yaghi@tsinghua.edu.cn
    \end{minipage}}
    
\date{}

\begin{document}
\twocolumn[
\maketitle
\begin{center}
    \large \textbf{Abstract}
\end{center}
\begin{center}
    \begin{minipage}{0.9\textwidth}
        \normalsize
In covalent organic frameworks (COFs), the geometry of the building units
predetermines the structure of the framework, yet comparable design principles relating
the COF linkers to the local motions that govern guest adsorption and diffusion as
well as mechanical response have largely remained beyond reach. Here, we combine synchrotron X-ray pair distribution function (PDF) analysis
with molecular dynamics (MD) simulations accelerated by machine learning
interatomic potentials to resolve and compare the local dynamics of two
three-dimensional imine-linked COFs, deliberately chosen to differ in the
geometry and rigidity of one of their two building units. Both COFs are constructed
from the same 4-connected square planar linker,
$4,4',4'',4'''$-(pyrene-1,3,6,8-tetrayl)tetraaniline (PyTTA), reticulated in
one case with the 12-connected nanographene
hexakis{[3,5-bis($p$-formylphenyl)-4,6-dimethoxyphenyl]}hexabenzocoronene
(HBC-LA$_{12}$) to give COF-412,
{[(HBC-LA$_{12}$)(PyTTA)$_3$]}$_{\mathit{imine}}$, and in the other with the 6-connected
triptycene 2,3,6,7,14,15-hexakis(4-formylphenyl)triptycene (Trip-CHO) to give
COF-564, {[(Trip-CHO)$_2$(PyTTA)$_3$]}$_{\mathit{imine}}$. The combined PDF-MD approach resolves the amplitude, the time scale, and the
spatial extent of the motion executed by all COF linkers, separating the
limited libration of the fused aromatic cores from the larger reorientational
tilts of the pendant aryl rings they carry and from the translation of whole
linkers within the net. The simulations further reveal that exchanging the
nanographene- for the triptycene-based linker tunes the equilibrium tilt and the positional
probability density of PyTTA, raising the tilt amplitude of its linker partner, and, notably, strengthens the coupling between the
translations of neighbouring linkers. This approach provides a blueprint
for understanding local dynamical properties in COFs, and shows them to
be selectable through the choice of building unit and amenable to design.

  \end{minipage}
\end{center}
]

\clearpage

\section*{Introduction}

\begin{figure*}[h!]
\centering
\includegraphics[width=\linewidth]{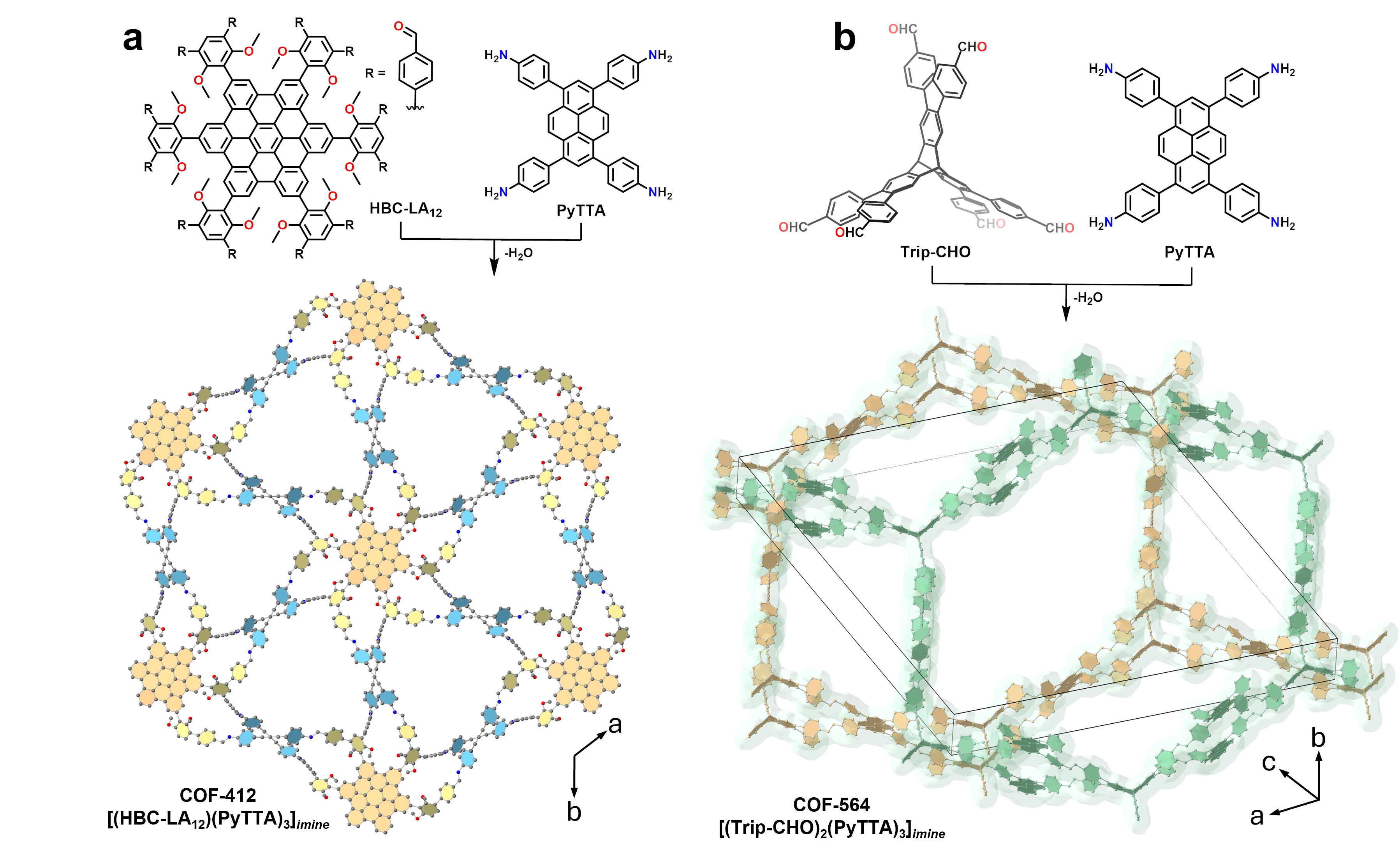}
\caption{Synthesis of COF-412 and COF-564. a) Condensation of the
twelve-connected, dodeca-benzaldehyde-functionalized nanographene
HBC-LA$_{12}$ with the four-connected tetraamine PyTTA to obtain COF-412.
b) Condensation of the six-connected aldehyde Trip-CHO with the four-connected
tetraamine PyTTA to obtain COF-564. Hydrogen atoms are omitted
in the crystal structures for clarity.}
\label{fig:fig1}
\end{figure*}

Covalent organic frameworks (COFs) are a class of crystalline porous materials
whose components are selected from a wide range of organic linkers and
covalently joined into predetermined topologies, affording stable backbones and
pore environments that are tunable in functionality and
metrics.\cite{Cote,COF2,diercks} The diversity of chemical building blocks and linkage
geometries makes COFs an object of design for structural and functional
control, with applications ranging from the selective capture of carbon dioxide \cite{cof_co2_n1,lyu,ALENAZI2025100365,10.1021/acs.iecr.5c03598}
and water \cite{COF_water_n2, 10.1039/d5ew00643k,sun_angew} from air, to gas separation \cite{ALOM2026125620,COF_sep,keskin_acs}, chemical sensing \cite{yue_chem_eur_j,10.1039/c9cs00299e}, and heterogeneous
catalysis \cite{Guo, Shirani_jmaterchemA}. However, this control has been exercised
almost entirely over the global static structures of COFs and has yet to be
extended to the rational design of their local dynamic properties, which, as in all
reticular materials, govern guest adsorption and diffusion as well as
mechanical response\cite{priyabrata_dalton,CCS}. That such motion pervades COF materials
is well documented. For instance, it has been shown that single crystals of \textbf{dia} COFs expand and contract
reversibly as guests enter and leave their pores \cite{Yao2025}, COFs may adapt their pore geometry to the guests they include \cite{Liu2024}, imine
linkages undergo pedal motion that influences framework flexibility \cite{Chi2023},
and aromatic rotors mounted within porous organic frameworks spin at megahertz
rates even at cryogenic temperatures \cite{Comotti2014}. \\
However, while reticular chemistry rests on the precise translation of molecular
geometry into extended frameworks, no defined principles yet exist by which local
structure may be translated into framework dynamics. With such principles in
hand, COFs could desirably be made locally rigid to maintain pore shape and
selectivity, or locally mobile to modulate the pore aperture, adapt to incoming
guests, and facilitate their diffusion. Establishing such design guidelines requires an
experimental probe sensitive both to the local and the extended order of
COFs, information beyond the reach of powder X-ray diffraction (PXRD) methods ordinarily
used to characterize COF systems \cite{MOF3_f,preprint_saber}. \\
We reasoned that the atomic pair
distribution function (PDF) obtained from X-ray total scattering experiments could meet
this need, for it converts the scattering of a material, Bragg and diffuse
alike, into a real space map of interatomic distances spanning local,
intermediate, and long range order in any solid regardless of
crystallinity \cite{terban_chem_rev,billings_pdf_review,Griffiths2024_NatMater,Klove2023_AdvMater,goodwin_n1,Zhu2021_AdvSci,cc_platero,Cliffe2014_NatCommun,RomeroMuniz2024_CSR,Bennett2014_AccChemRes,Bennett2010_PRL,Cao2012_ChemCommun,Gaillac2017_NatMater,Keen2018_PCCP,Terban2018_Nanoscale,Molina2021_MicroMeso,Sapnik2022_NatCommun,Sapnik2023_CommunChem}. In conjunction with theoretical methods such as molecular dynamics (MD)
simulations, now extended by machine learning interatomic potentials (MLIPs) to
length and time scales beyond the reach of \emph{ab initio} approaches \cite{wood2026umafamilyuniversalmodels}, the
features of the PDF may further be traced quantitatively to the interatomic
vectors that produce them and linked to the underlying dynamical behaviour \cite{tavjacscof}. \\ 
Here, we combine synchrotron X-ray PDF analysis with MLIP-accelerated MD
simulations to resolve the local dynamics of selected three-dimensional (3D) COFs and
point toward the design of COFs with tunable local dynamics.
We demonstrate this approach on two 3D imine-linked COFs built
from the same 4-connected square planar linker,
$4,4',4'',4'''$-(pyrene-1,3,6,8-tetrayl)tetraaniline (PyTTA), whose fused
pyrene core is planar and rigid. In COF-412,
[(HBC-LA$_{12}$)(PyTTA)$_3$]$_{\mathit{imine}}$, PyTTA is reticulated with the
12-connected hexagonal prismatic nanographene
hexakis[3,5-bis($p$-formylphenyl)-4,6-dimethoxyphenyl]hexabenzocoronene
(HBC-LA$_{12}$), whose hexabenzocoronene core is likewise a rigid, planar sheet
of fused rings and whose twelve connecting arms are carried on dimethoxyphenyl
spacers, affording
the (3,4,6)-connected \textbf{cez} topology and the highest porosity reported
for COFs to date (Figure \ref{fig:fig1}a).\cite{MOF3_f} Both building units of COF-412 therefore
present extended aromatic surfaces of comparable rigidity. In COF-564,
[(Trip-CHO)$_2$(PyTTA)$_3$]$_{\mathit{imine}}$, that rigidity is modified while
PyTTA is left in place, the nanographene being replaced by the 6-connected
trigonal prismatic
2,3,6,7,14,15-hexakis(4-formylphenyl)triptycene (Trip-CHO), whose three arene peripheral groups
are fixed at $120^{\circ}$ about a bridgehead axis, so that an extended planar
surface gives way to a compact unit of 3D shape and the
framework adopts the (4,6)-connected \textbf{stp} topology (Figure \ref{fig:fig1}b) \cite{arxiv_trypticene}. Holding PyTTA constant while varying its linker partner isolates the influence of the
second building unit on framework motion. With the ML-MD models of both COFs
validated against the measured PDFs, we analyze the MD trajectories to resolve
the local structure of each framework and to quantify the reorientational and
translational motion of its linkers. We find that replacing the nanographene-based linker by
the triptycene-based building unit in COF-564 influences both the tilt angle and the positional
disorder of PyTTA and, importantly, allows the translations of neighbouring
linkers to become correlated, a coupling that is impeded in COF-412. Taken
together, these results establish how the local dynamics of COFs connect to the structural features encoded in the PDF, and how the dynamics of their
building units may be understood and directed.

\section*{Results and Discussion}

\begin{figure*}[h!]
\centering
\includegraphics[width=\linewidth]{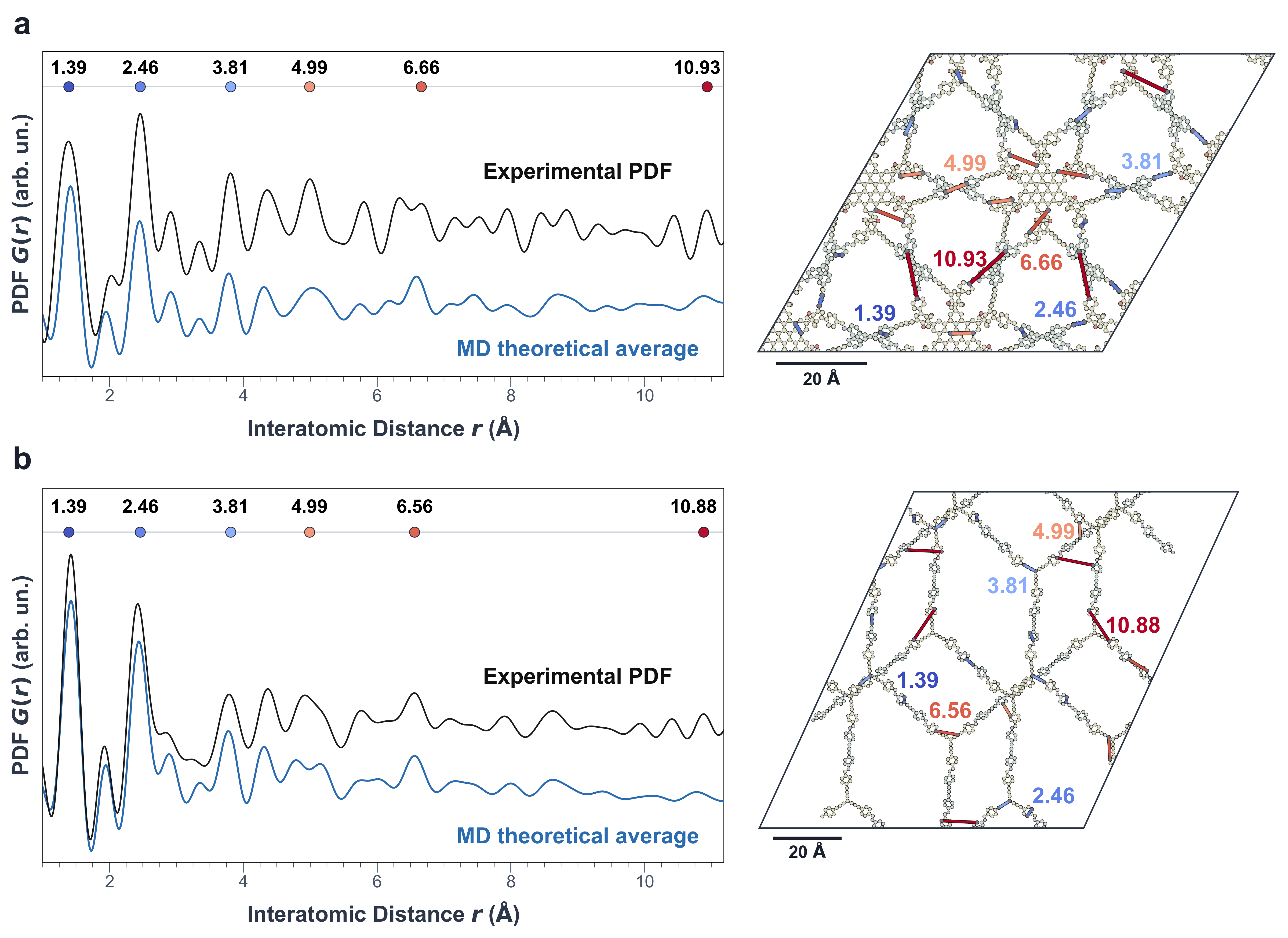}
\caption{Experimental and MD-averaged theoretical PDFs of a) COF-412
and b) COF-564. The measured PDFs (black) and the PDFs computed as an
ensemble average over the MLIP-MD trajectory (blue) are displayed in the left panels. The
markers above each panel indicate the positions of selected maxima in the experimental PDFs. A visualization of the COF structures where these maxima are assigned to interatomic vectors drawn on the corresponding unit cells is displayed in the right panels. Hydrogen atoms are omitted for clarity.}
\label{fig:fig2}
\end{figure*}

Both COFs were obtained by imine condensation, and their syntheses are summarized in
Figure~\ref{fig:fig1}. COF-412 was prepared by condensing the
dodeca-benzaldehyde-functionalized nanographene HBC-LA$_{12}$ with PyTTA
(Figure~\ref{fig:fig1}a), and COF-564 by condensing the hexa-benzaldehyde
triptycene Trip-CHO with the same tetraamine (Figure~\ref{fig:fig1}b). By virtue of the differing geometry and connectivity of the polyaldehyde linkers, the two COFs
adopt distinct topologies, as shown in Figure~\ref{fig:fig1}. In COF-412 the twelve arms of each nanographene linker extend within a
hexagonal prismatic envelope and are met by square planar PyTTA units, leading to
a \textbf{cez} net. In COF-564 the six arms of each triptycene linker project from a
bridgehead axis into three directions, resulting in a \textbf{stp} net \cite{arxiv_trypticene}. Additional details on the synthetic
procedures and related characterization of the COFs are provided in Section S1 of the SI (see Figures S1--S4).

The local structure of COFs is challenging to access experimentally \cite{lotsch_chemsci,RomeroMuniz2021_ACSAMI}. Their synthesis often yields
polycrystalline rather than single crystalline products, so that their
structural characterization has been largely limited to structures modeled from PXRD data, which are poorly sensitive to the positional disorder of the
linkers and the stacking of their layers. Moreover, the COF frameworks are composed entirely of light elements (C, N, O, B, H) and offer no absorbing center to hard X-ray spectroscopic probes, such as hard X-ray
absorption spectroscopy (XAS) \cite{tav_dalton,tavani_jpcl_Zn,tav_pccp,delgiudice,tav_acs}, that serve the study of local properties in metal-organic
frameworks \cite{catal13101338,MOF1f, MOF2f}.
To obtain accurate local structure information, we collected X-ray PDF total
scattering data on COF-412 and COF-564 at
beamline 28-ID-1 of the National Synchrotron Light Source II at 290~K using
radiation of wavelength $\lambda = 0.1665$~\AA{}. The resulting PDFs, shown as
black curves in Figure~\ref{fig:fig2}a,b, display a series of well-defined peaks
that persist above 12~\AA{}, indicating that both frameworks retain
short and intermediate range order.

To interpret these structural correlations, we computed the corresponding PDFs
as ensemble averages over MD simulations of each COF, which made use of the
universal model for atoms (UMA)
\cite{wood2026umafamilyuniversalmodels} (see the Supporting Information,
Section S2, for additional details on the employed MD procedure). UMA is a transferable MLIP trained across
molecular and periodic quantum-mechanical datasets of wide chemical diversity,
including organic molecules, molecular crystals, and reticular materials. It describes
at near-DFT accuracy both the covalent bonding within the linkers and their
non-covalent contacts, the two interactions that together set the
local structure of an imine-linked COF, while granting access to length and
time scales beyond the reach of \emph{ab initio} methods when validated against high-fidelity experimental data such as synchrotron PDF measurements \cite{wood2026umafamilyuniversalmodels,tavjacscof}. In particular, 100 MD snapshots were extracted at equal intervals along each
trajectory for both COFs, and the ensemble PDFs were computed as their converged average
(Figures~S5--S6). As displayed in Figures~\ref{fig:fig2}a,b, the MD-averaged PDFs agree with the
experimental data of both COFs in the positions of the main peaks, from the
intramolecular bonds at low $r$ to the inter-linker correlations beyond
9~\AA{}. The agreement extends to features
that are diagnostic of the local geometry of the building units. For instance,
the relative heights of the maxima at 1.39 and 2.46~\AA{}, which measure the
balance of nearest- and second-neighbour aromatic contacts, are reproduced for
both COFs, as the intra-linker correlation at 3.81~\AA{} and the
widths of these peaks, which report on the spread of conformations sampled by
the linkers. Above about 9~\AA{} the measured PDFs display oscillations that decay weakly with $r$. For COF-564 the computed PDF reproduces the positions of these maxima, though with reduced amplitudes (Figure~\ref{fig:fig2}b), whereas for COF-412
the computed peaks at these distances are less intense than the measured
ones and their positions are slightly overestimated by the UMA-based MD model (Figure~\ref{fig:fig2}a) \cite{tavjacscof}.

We next employed the MD models to assign the individual PDF peaks to specific
interatomic vectors, mapping them onto the respective unit cells shown in the
right panels of Figure~\ref{fig:fig2}. Assignments were made by counting bonds
along the shortest covalent path in the structural model of each COF
(Tables S1-S2). Below about 4.5~\AA{} the PDFs of the two COFs are very similar, both
being dominated by intra-linker correlations. The two sharp maxima at 1.39 and
2.46~\AA{} arise from the nearest-neighbour aromatic C--C bond and from the
second-neighbour (\textit{meta}) C$\cdots$C distance across the aromatic rings,
as expected of materials built from phenylene and fused-ring units. The maximum
at 2.89~\AA{} arises from the \textit{para} separation across a single
six-membered ring and is more intense in COF-412, consistent with the larger
fraction of fused rings in its linkers. The maximum at 3.81~\AA{} is assigned
to a three-bond path crossing a ring junction into the adjacent ring. The
maxima between 4.3 and 6~\AA{} arise from four- to seven-bond paths that walk
outward along a single linker for both COFs. In COF-564, the peak centered around 4.99~\AA{} is split into a main peak and a shoulder at higher $r$, both of which can be assigned through the employed theoretical model. As shown in  Figure~\ref{fig:fig2}b, the MD-averaged PDF of COF-564 displays a maximum
at 4.79~\AA{} which arises from four-bond C$\cdots$C paths and is dominated by pairs
connecting the pyrene core of PyTTA to its phenyl arms (41\%) and the
triptycene blades of Trip-CHO to their pendant aryl rings (31\%). Conversely, the PDF shoulder peak at 5.14~\AA{} arises from the
corresponding five-bond paths, with nearly equal contributions from the
blade--aryl pairs of Trip-CHO (34\%) and the core--arm pairs of PyTTA (33\%). The PDFs of the two COFs differ more markedly above 6~\AA{}. The
maximum at 6.66~\AA{} in COF-412 arises mainly from six- to nine-bond paths
within its larger nanographene linker, whereas the corresponding feature of
COF-564 lies 0.10~\AA{} lower, at 6.56~\AA{}, and receives contribution from
six-bond paths within PyTTA together with pairs bridging the imine linkage
into the aryl rings of Trip-CHO. Beyond this distance, inter-linker
correlations increasingly contribute to the PDFs of both frameworks. The maxima at
approximately 7.95 and 10.9~\AA{} are present in both PDFs and receive
contributions from pairs that either bridge the imine linkage or span two
linkers that are not directly connected (see Figure~\ref{fig:fig2}, right
panels).

\begin{figure*}[h!]
\centering
\includegraphics[width=0.85\linewidth]{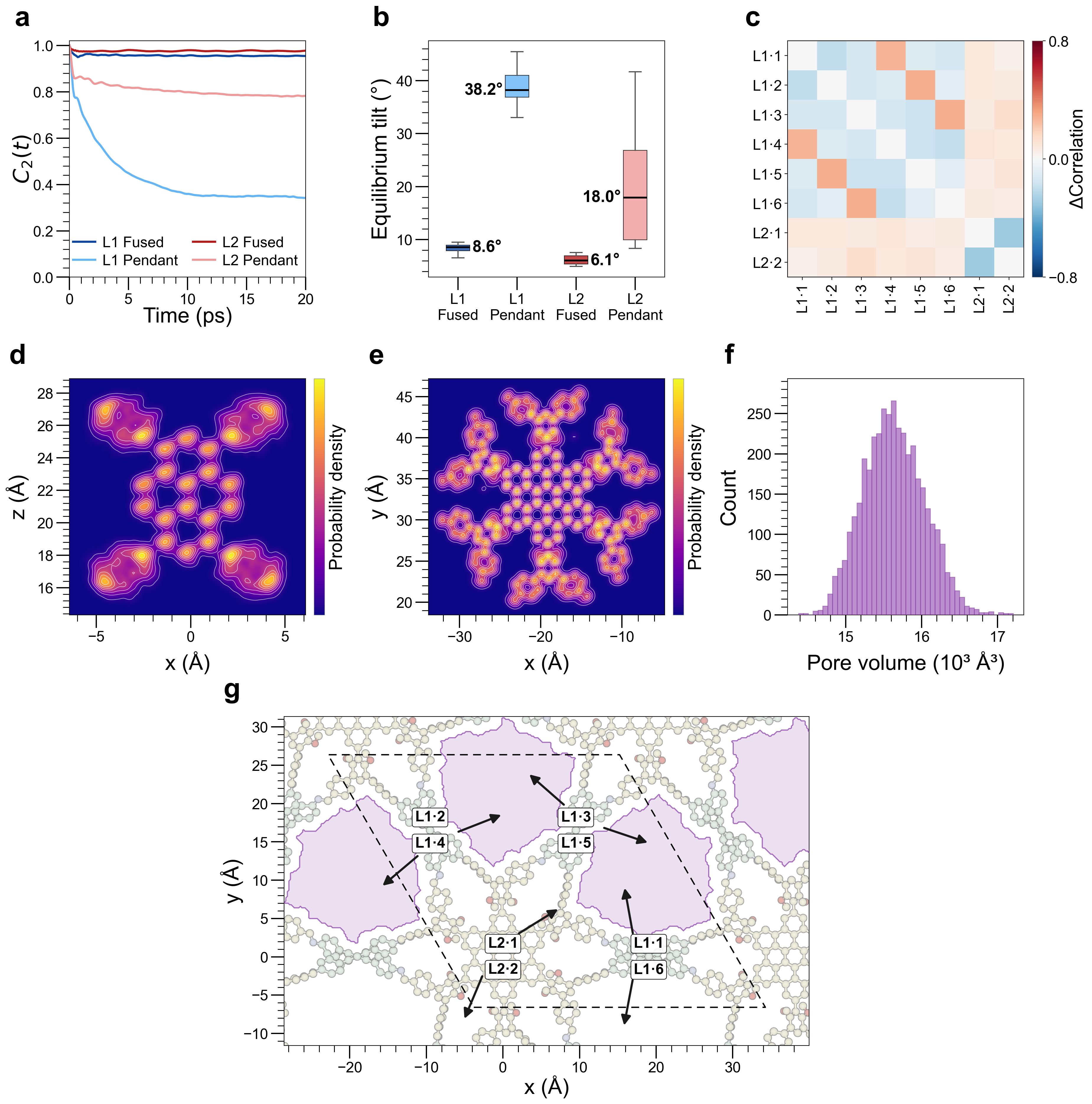}
\caption{Local dynamics of COF-412 obtained from the analysis of MLIP-MD trajectories. a) Second-order orientational
correlation function $C_2(t)$ of the ring normal of the fused cores and pendant aryl rings of the
PyTTA (L1) and HBC-LA$_{12}$ (L2) linkers. b) Distributions of the equilibrium
tilt angle of the same four groups, with mean values indicated. c)
Cross-correlation matrix of the centre-of-mass displacements of the six
crystallographically distinct PyTTA units (L1-1 to L1-6) and the two
HBC-LA$_{12}$ units (L2-1 and L2-2). d,e) Positional probability densities of selected PyTTA (d) and HBC-LA$_{12}$ (e) linkers. f) Distribution of the pore volume sampled over the MD trajectory.
g) Visualization of the directions of motion of the linkers within the COF cell, where selected pore areas are highlighted.}
\label{fig:fig3}
\end{figure*}

Having established that the MD simulations reproduce the short-range order
of both COFs and the intermediate-range order of COF-564, we
turned to investigating the dynamic properties of the COFs encoded in the MD trajectories, beginning with
COF-412 (Figure~\ref{fig:fig3}). We treated each linker in two parts, the fused
aromatic core and the covalently linked pendant aryl rings, and monitored the
reorientation of each through the second-order orientational correlation
function $C_2(t) = \langle P_2(\mathbf{n}(t_0)\cdot\mathbf{n}(t_0+t))\rangle$, where
$P_2$ is the second Legendre polynomial, $\mathbf{n}$ is the unit normal of
the aromatic plane, and the average runs over time origins
$t_0$~\cite{MOF2f,MOF1f} (Figure~\ref{fig:fig3}a). The behaviour of the two fused
cores is similar. Both the pyrene core of PyTTA (L1) and the
hexabenzocoronene core of HBC-LA$_{12}$ (L2) exhibit $C_2(t)$ values near unity
throughout a 20~ps MD simulation time window, indicating that the fused cores of neither linker reorient on this
time scale. The $C_2(t)$ values of the linker pendant rings, by contrast, decay markedly and to different
extents, so that within COF-412 the two linkers
carry pendant groups of substantially different mobility.

The amplitudes of these motions follow the same ordering
(Figure~\ref{fig:fig3}b). The tilt angle of the fused cores is confined
to median values of $8.6^{\circ}$ for L1 and $6.1^{\circ}$ for L2, whereas the pendant rings
exhibit ring equilibrium tilt median values of $38.2^{\circ}$ and $18.0^{\circ}$, respectively. These analyses indicate that the fused cores librate about their equilibrium
orientation while the pendant rings undergo large-amplitude reorientation.
COF-412 may therefore be described as rigid in its linker cores and mobile at
the joints that bridge PyTTA and HBC-LA$_{12}$.

To determine whether the building units of COF-412 translate independently or
in concert, we computed the cross-correlation matrix of the linker centres of
mass (Figure~\ref{fig:fig3}c). Its normalised elements,
$C_{ij} = \langle \Delta\mathbf{R}_i \cdot \Delta\mathbf{R}_j \rangle /
(\langle |\Delta\mathbf{R}_i|^2\rangle \langle
|\Delta\mathbf{R}_j|^2\rangle)^{1/2}$, where $\Delta\mathbf{R}_i$ is the
displacement of the centre of mass of linker $i$ from its mean position, equal
unity for two linkers that move together and minus one for two linkers whose
displacements are opposed. Because the conserved centre of mass of the
simulation cell imposes a weak negative correlation on all pairs even for
fully independent motion, the matrix is reported relative to this baseline,
so that the displayed values vanish for uncoupled linkers (see Section~S2 of
the SI). The off-diagonal elements among the six PyTTA units
(L1-1 to L1-6) and the two nanographene units (L2-1 and L2-2) are
comparatively weak, indicating that the linkers of COF-412 translate largely
independently of one another. Closer inspection nonetheless reveals that the displacements of the two nanographene
units (L2-1 and L2-2) are on average opposed, as they exhibit the strongest
off-diagonal element of the matrix (see darker blue regions in
Figure~\ref{fig:fig3}c). We attribute this anticorrelation to the pore the two linkers share, for they
face each other across a single aperture, so that the approach of one toward
the pore centre is balanced by a displacement of the other
(Figure~\ref{fig:fig3}g).

This dynamic behaviour affects the positional probability
densities of both COF linkers, which are shown in Figures~\ref{fig:fig3}d,e. In both probability density projections the fused cores appear
as sharp, atomically resolved features, whereas the pendant rings are smeared
into broad lobes, reflecting the greater reorientational freedom of the pendant aryl groups \cite{tavjacscof}. Despite this local mobility, the pores of COF-412 exhibit a narrowly
distributed volume, as shown in Figure~\ref{fig:fig3}f which displays the MD-derived pore volume analysis (refer to Section~S2 of the SI for additional details). The width of the
resulting distribution amounts to a few percent of the mean volume, indicating
that, within the model, the large local librations of the linker pendant aromatic rings occur
without significantly altering the metrics of the pore they line. We note that these values refer to the simulated COF, so
that the absolute pore volume of COF-412 may carry a corresponding offset.

\begin{figure*}[h!]
\centering
\includegraphics[width=0.85\linewidth]{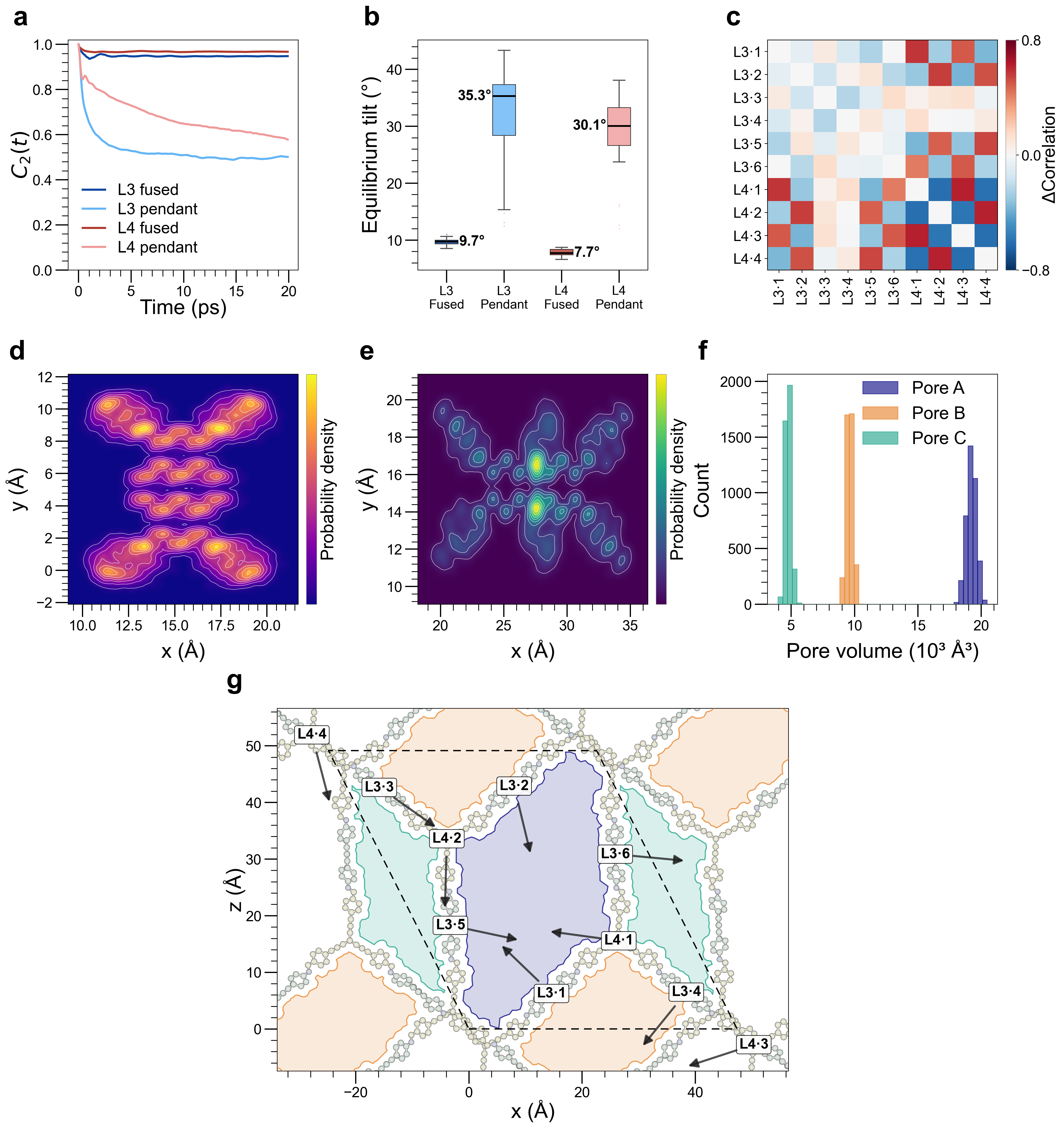}
\caption{Local dynamics of COF-564 obtained from the analysis of MLIP-MD trajectories. a) Second-order orientational
correlation function $C_2(t)$ of the ring normal of the fused cores and pendant aryl rings of the
PyTTA (L1) and Trip-CHO (L4) linkers. b) Distributions of the equilibrium
tilt angle of the same four groups, with mean values indicated. c)
Cross-correlation matrix of the centre-of-mass displacements of the six
crystallographically distinct PyTTA units (L3-1 to L3-6) and the four
Trip-CHO units (L4-1 and L4-2). d,e) Positional probability densities of selected PyTTA (d) and Trip-CHO (e) linkers. f) Distribution of the three pore volumes sampled over the MD trajectory.
g) Visualization of the directions of motion of the linkers within the COF cell, where selected pore areas are highlighted.}
\label{fig:fig4}
\end{figure*}

To determine how the second building unit influences the COF local dynamics, we
applied the same analysis to COF-564, in which PyTTA is retained while the
extended planar nanographene linker is replaced by the compact
triptycene-based building unit, so that any difference in the dynamics may be
attributed to this substitution. In triptycene, the three arene wings are held
about a bridgehead axis rather than fused into a single plane as in the
nanographene. We reasoned that this exchange would replace a broad aromatic
platform, which anchors twelve closely spaced arms to one rigid sheet whose
plane defines their equilibrium orientation, by a compact node bearing six
arms, so that each arm would be less crowded by its neighbours and less
constrained by the surface on which it is mounted, while the building unit
itself remained rigid and directionally defined. The pendant aryl arms of the
partner should therefore sample larger tilt amplitudes than those borne on the
nanographene sheet, and the motion of PyTTA, although the linker itself is
unchanged, should be altered at least in part.

As shown in Figure~\ref{fig:fig4}, the pattern of motion found in COF-564 exhibits differences compared to that of COF-412. The fused
cores of both linkers again show almost no reorientation, with the $C_2(t)$ values of the pyrene core of
PyTTA (L3) and of the triptycene core of Trip-CHO (L4) remaining above 0.95
over 20~ps, and their pendant rings decaying to approximately 0.50 and 0.58,
respectively (Figure~\ref{fig:fig4}a). The corresponding equilibrium tilt amplitudes display median values of
$9.7^{\circ}$ and $7.7^{\circ}$ for the fused cores of PyTTA and Trip-CHO, and of
$35.3^{\circ}$ and $30.1^{\circ}$ for their pendant rings
(Figure~\ref{fig:fig4}b). Notably, the mobility of PyTTA is itself altered by
the substitution as the tilt of its fused core rises by $1.1^{\circ}$ from the median value of
$8.6^{\circ}$ found in COF-412 while that of its pendant rings falls by
$2.9^{\circ}$ from a median of $38.2^{\circ}$. Although small, these shifts are in line with our hypothesis that a building unit retains only part of its
dynamic signature when its linker partner is changed. We
attribute the freer libration of the pyrene sheet to the loss of the rigid and broad
aromatic platform of HBC-LA$_{12}$ opposite it. The mobility of the Trip-CHO linker partner
changes more significantly. The tilt of the pendant rings of the trypticene-based linker rises by $12.1^{\circ}$, from
$18.0^{\circ}$ for the arms of the nanographene to $30.1^{\circ}$ for those of
the triptycene, and that of its fused core by $1.6^{\circ}$, from
$6.1^{\circ}$ to $7.7^{\circ}$, as anticipated for arms borne on a compact node
rather than closely spaced about one broad sheet. Together, these changes bring
the pendant rings of the two linkers to comparable mobility, differing by
$5.2^{\circ}$ in COF-564, whereas in COF-412 they differ by $20.2^{\circ}$.
Part of the dynamic signature of a building unit is therefore carried with it
from one COF into another, and part is set by the partner with which it
is reticulated.

The translational behaviour changes more markedly. The cross-correlation
matrix of COF-564 (Figure~\ref{fig:fig4}c) exhibits appreciable off-diagonal
elements among the six PyTTA units (L3-1 to L3-6) and the four triptycene units
(L4-1 to L4-4). In particular, the PyTTA--PyTTA
elements remain weak and mildly anticorrelated, much as in COF-412. Interestingly, several pairs among the triptycene units are markedly negatively and positively correlated, while also the
PyTTA--triptycene matrix elements display correlations. This pattern may be understood by extending the picture drawn for
the \textbf{cez} topology: triptycene-based linkers that bound a common pore displace in mutually opposed
fashion as one approaches the pore centre, as found for the two nanographene
cores in COF-412,
whereas triptycene-based units on the same side of a pore are carried along
together, indicating that the motions of neighbouring linkers in the
COF-564 are coupled to a degree not seen in COF-412. The
probability densities of L3 and L4 are shown in Figure~\ref{fig:fig4}d,e and localize the fused
cores and delocalize the pendant rings of both linker.

Where COF-412
exhibits one main kind of pore (see Figure 3g, purple regions), the six-connected
triptycene node of COF-564 results in three kinds of pores, of approximately
$19$, $9.7$, and $4.8 \times 10^3$~\AA{}$^3$, each with a narrow distribution
of volume (Figure~\ref{fig:fig4}f), which tile the lattice in a repeating
pattern (Figure~\ref{fig:fig4}g). The exchange of one building unit therefore
alters not only the amplitude of local motion but also the number and size of
the pores whose walls that motion modulates. The narrow spread of all
pore volume distributions indicates retention of the pore metrics in both COFs,
so that local mobility is accommodated without pore deformation.

\section*{Conclusion}

In summary, we have combined synchrotron X-ray PDF analysis with MD simulations
accelerated by MLIPs to resolve the local
structure and dynamics of two 3D imine-linked COFs built from a common linker.
The combined PDF measurements and MD simulations establish a consistent picture of
motion. Fused aromatic cores librate within median values of $10^{\circ}$ at 290 K and show reduced
reorientational ability, while the pendant aryl rings they carry sample tilt angles three
to more than four times larger. Exchanging the nanographene-based HBC-LA$_{12}$
linker of COF-412 for the triptycene-based Trip-CHO linker of COF-564 (i) influences
the dynamics of the shared PyTTA linker, (ii) raises the tilt
amplitude of the pendant rings of the Trip-CHO partner by $12.1^{\circ}$ at 290 K, and (iii) increases
the coupling between the translational motions of neighbouring linkers. Two comparative
design principles follow. First, stiffness and mobility occupy separate and
predictable positions in these frameworks, rigidity residing in the fused cores
and motion at the pendant rings and at the joints between building units, so
that each can be placed by the choice of linker without compromising the pore
metrics. Second, the mobility of a linker is a property of the reticulated pair
rather than of the linker alone. Indeed, a building unit carries only part of its
dynamic signature from one COF to the other, so that pairing a given linker
with a broad aromatic sheet or with a compact node offers a direct handle on
its dynamical properties.\\
We anticipate that the employed method will aid the screening of
candidate combinations of COF building units before synthesis, with the
measured PDFs serving as the experimental benchmark for theoretical validation.
Extension of this approach to guest-loaded COFs would then allow the local
dynamics to be followed in realistic operating conditions, and
COFs to be designed in which the intra- and inter-linker dynamics
are tailored to function.

\section*{Data Availability}
Experimental PDF data will be made available on the Zenodo platform.

\section*{Supporting Information}

General experimental methods and materials; syntheses and characterization of
COF-412 and COF-564; total scattering data collection and PDF processing;
MLIP-MD protocol and convergence tests; additional PDF results.

\section*{Acknowledgments}
F.T. acknowledges funding from the European Union’s Horizon Europe research and innovation program under the Marie Skłodowska-Curie grant agreement No. 101201413 (AQUAFRAME). Effort of S.J.L.B. and C.M. on PDF analysis, modeling and writing was supported by U.S. Department of Energy, Office of Science, Office of Basic Energy Sciences (DOE-BES) under contract No. DESC0024141. This research used beamline 28-ID-1 of the National Synchrotron Light Source II, a U.S. DOE Office of Science User Facility operated for the DOE Office of Science by Brookhaven National Laboratory under Contract No. DE-SC0012704. Claude-Opus5.0 was used for grammar and sentence structure checking during the review and editing of this manuscript.

\providecommand{\latin}[1]{#1}
\makeatletter
\providecommand{\doi}
  {\begingroup\let\do\@makeother\dospecials
  \catcode`\{=1 \catcode`\}=2 \doi@aux}
\providecommand{\doi@aux}[1]{\endgroup\texttt{#1}}
\makeatother
\providecommand*\mcitethebibliography{\thebibliography}
\csname @ifundefined\endcsname{endmcitethebibliography}
  {\let\endmcitethebibliography\endthebibliography}{}

\end{document}